\documentclass[prl,twocolumn,superscriptaddress,showpacs,amsmath,amssymb,floatfix]{revtex4-1}
\usepackage{graphicx,color,epsfig,bm,float}% Include figure files

\begin{document}

\title{Time periodicity and dynamical stability in two-boson systems}

\author{Jose Reslen}

\affiliation{Coordinaci\'on de F\'{\i}sica, Universidad del Atl\'antico, Km~7 Antigua v\'{\i}a a Puerto Colombia, A.A. 1890, Barranquilla, Colombia.}

\date{\today}

\begin{abstract}
We calculate the period of recurrence of dynamical systems comprising two 
interacting bosons. A number of theoretical issues 
related to this problem are discussed, in particular, the conditions for
small periodicity. The knowledge gathered in this way is then
used to propose a notion of dynamical stability based on the stability
of the period. Dynamical simulations show good agreement with the
proposed scheme. We also apply the results to the phenomenon known
as coherent population trapping and find stability conditions
in this specific case.
\end{abstract}

\maketitle

\section{Introduction}
An important result known as the Recurrence Theorem
\cite{Barreira,Shepe} establishes that quantum as well
as classical systems will come very close to their initial state
at some time during their evolution. This time is known as
the recurrence time, and the theorem applies in the context of
closed systems. This kind of recurrence behavior
has been observed experimentally in quantum 
systems such as Rydberg states of the Hydrogen atom \cite{Yeazell}. A
straightforward consequence of the theorem is that closed systems
possess, at least in a broad sense, an intrinsic or natural frequency
given by the inverse of the recurrence time. 

The intrinsic frequency characterizes the response of the system to
external driving. The amplitude of the response is
maximally enhanced when the driving frequency matches the natural
frequency of the system~\cite{Reslen1}.  This enhancement underlies
a number of important physical phenomena and its understanding
is of fundamental interest.  Assuming that the driving frequency
is constant, one should minimize fluctuations of the natural
frequency in order to amplify the response, especially when the
parameters of the system are subject to small perturbations. 

Certain classical aspects of stability in two-body systems have been
discussed in \cite{Politi}.  Likewise, the stability of
quantum dynamics as a result of a small change in the parameters has
been discussed in several works, for example in Ref.~\cite{Prozen} or
Ref.~\cite{Peres}.  In models with classical counterparts,
the inner product between quantum states of two systems with the same initial
condition but with slightly different parameters remains close to unity 
during the evolution as
long as the initial wave function is well localized inside a
stable island of the classical map. The opposite takes place when the
initial condition is localized in a chaotic region. Additionally, recurrences
in bosonic systems have been studied in Ref.~\cite{Donovov} as 
a form of state transfer in the time domain. Our proposal is different
in that we study Hamiltonians displaying two-body interaction.

Here we intend to approach the issue of stability by studying the
behavior of periodicity in a quantum model.  The study contains a
moderate analysis of the period, which lays the foundation for the
subsequent argument concerning stability. In the absence of a
classical analogue, we base our approach exclusively on the
eigenenergies of the Hamiltonian, making little reference to the
initial state. 

In this paper we focus initially on a series of issues related to the 
recurrence period of a two-boson system,
particularly, the question of how small the period can be as a
function of the parameters.  Similarly, we find the set of parameters
and the directions along which such parameters must be tuned in order
to keep the period constant. Finally, we suggest an application to a
quantum optical technique known as coherent population trapping. The
results we obtain are also relevant in other scenarios. For instance, 
in quantum computation, where some information protocols~\cite{Lisi} or
quantum gates are subject to perturbations of the parameters. 
Additionally, this study gives insight into the physics
of few-boson systems~\cite{Cao,Quiroga,Blas}, which constitute the
basis of more complex structures.

In the language of second-quantization, the quantum state is written with
reference to occupation modes of unperturbed levels.  In this context,
let us focus on a model featuring two-body interaction, such as
\begin{equation}
\hat{H} =  J(\hat{a}_1^{\dagger} \hat{a}_2 + \hat{a}_2^{\dagger} \hat{a}_1) +  \sum_{k=1}^2 \frac{U_k}{2} \hat{a}_k^{\dagger} \hat{a}_k (\hat{a}_k^{\dagger} \hat{a}_k - 1) + \epsilon_k \hat{a}_k^{\dagger} \hat{a}_k   . 
\label{g1_1}
\end{equation}

As usual, the exchange term $J$ and the unperturbed energies
$\epsilon_1$ and $\epsilon_2$ define the single-body response, while
$U_1$ and $U_2$ determine the intensity of the interaction among
particles and can be seen as a nonlinear contribution. The mode
operators satisfy the usual bosonic relations
$[\hat{a}_1,\hat{a}_1^{\dagger}]=[\hat{a}_2,\hat{a}_2^{\dagger}]=1$,
etc. The unperturbed system can be probed by looking at the absorption
profile of an incident laser of frequency $\nu =
|\epsilon_2-\epsilon_1|/\hbar$. The total number of particles

\begin{equation}
M = \hat{a}_1^{\dagger} \hat{a}_1 +  \hat{a}_2^{\dagger} \hat{a}_2,
\label{g1_2}
\end{equation}
is a conserved quantity. The proposed system reduces to a two-level
model when $M=1$. Hamiltonian (\ref{g1_1}) can be rearranged into
the form
\begin{equation}
\hat{H} = \eta \hat{a}_1^{\dagger} \hat{a}_1 \hat{a}_1^{\dagger} \hat{a}_1 - \mu \hat{a}_1^{\dagger} \hat{a}_1 + \hat{a}_1^{\dagger} \hat{a}_2 + \hat{a}_2^{\dagger} \hat{a}_1,
\label{g1_3}
\end{equation}
where, for $M=2$, the parameters $\eta$, $\mu$ and $\Delta$ turn out to be
\begin{equation}
\eta = \frac{U_1 + U_2}{2 J}, \text { } \mu = \frac{3 U_2 + U_1}{2 J} + \frac{\Delta}{J}, \text{ } \Delta = \epsilon_2 - \epsilon_1.
\label{g1_4}
\end{equation}

In writing the previous identities we have chosen $J$ to be our energy
unit \cite{note2}. The state evolution is given by the expression (in what follows,
we set $\hbar=1$)
\begin{equation}
|\psi(t) \rangle = q_1 e^{-i E_1 t} |E_1 \rangle + q_2 e^{-i E_2 t} |E_2 \rangle + q_3 e^{-i E_3 t} |E_3 \rangle.
\label{g2_3}
\end{equation}

Hence, periodicity arises whenever \cite{Godsil}
\begin{equation}
E_1 T = 2\pi n_1, \text{ } E_2 T = 2\pi n_2, \text{ } E_3 T = 2\pi n_3.
\label{g2_4}
\end{equation}
where $T$ is the period of the recurrence, and $n_1$, $n_2$ and $n_3$ are
integer numbers.  In this case the periodicity is absolute as the quantum
state recurs identically at regular intervals and the corresponding
evolution operator equals the unity operator. Another form of
periodicity \cite{Ralph} emerges by considering instances in which the
quantum state recurs up to a phase, i.e.
\begin{equation}
|\psi(t + \tau)\rangle = e^{-i\phi}|\psi(t) \rangle.
\label{g2_7}
\end{equation}

We call this partial periodicity, since the phase factor may generate
quantum interference effects. As an example, let us look at Hamiltonian 
(\ref{g1_3}) for a single particle
\begin{equation}
\hat{H} = 
\left (
\begin{array}{cc}
-\mu & 1 \\
  1  & 0  
\end{array}
\right ).
\label{g2_2}
\end{equation}

Absolute periodicity occurs whenever the ratio of energies is a
fractional number
\begin{equation}
  x = \frac{E_1}{E_2} = \frac{\mu-\sqrt{\mu^2+4}}{\mu+\sqrt{\mu^2+4}} = \frac{n_1}{n_2}.
\label{g2_5}
\end{equation}

$T$ can be found from Eq. (\ref{g2_4})
\begin{equation}
T = 2 \pi n_2 \sqrt{-x} = 2 \pi \sqrt{- n_1 n_2}.
\label{g2_6}
\end{equation}

In Eq. (\ref{g2_5}) we can  define $E_1$ and $E_2$ so that 
$|n_2| \geqslant |n_1|$ and therefore $-1 \leqslant x <0$.
In principle, there is no limit on the maximum value of $T$. 
Conversely, the minimum value is $T=2\pi$ and takes place at $\mu=0$.

In a similar way, partial periodicity derives from

\begin{equation}
E_1 \tau = \phi, \text{ } E_2 \tau = \phi + 2\pi,
\label{g2_8}
\end{equation}

and therefore

\begin{equation}
\tau = \frac{2\pi}{|E_2-E_1|} = \frac{2 \pi}{\sqrt{\mu^2+4}} = 2\pi\sqrt{\frac{-x}{(1-x)^2 }}.
\label{g2_9}
\end{equation}

The first equality is consistent with the view that the natural
frequency of the system is proportional to the difference of its two
eigenenergies. Unlike $T$, $\tau$ reaches a maximum $\tau=\pi$ at
$\mu=0$ and goes down asymptotically to zero as $\mu\rightarrow \pm
\infty$, two limits in which one of the eigenenergies dominates the
spectrum and the Hamiltonian is almost singular.  This shows that
$\tau$ is maximum when $T$ is minimum and that $\tau$ goes to zero as
$T$ goes to infinity.  Two-level systems always display partial
periodicity, but not necessarily absolute periodicity.

\section{Two particles}
Let us now probe these periodicity concepts in a larger
system. For $M=2$, Hamiltonian~(\ref{g1_3}) takes the matrix form
\begin{equation}
\hat{H} = 
\left (
\begin{array}{ccc}
4\eta-2\mu & \sqrt{2} & 0 \\
\sqrt{2} & \eta - \mu & \sqrt{2} \\
0 & \sqrt{2} & 0
\end{array}
\right ).
\label{g1_5}
\end{equation}

The energies are the solutions of the characteristic equation
\begin{equation}
E^3 + \alpha E^2 + \beta E + \gamma = 0,
\label{g1_6}
\end{equation}
where
\begin{eqnarray}
& \alpha = -5 \eta + 3 \mu, \label{g1_7a} & \\  
& \beta = 2(2 \eta - \mu)(\eta - \mu) - 4, \label{g1_7b} & \\ 
\mathrm{and}\nonumber\\
& \gamma = 4 (2 \eta - \mu). \label{g1_7c}
\end{eqnarray}

From the previous equalities we can see that if an energy $E$
is a solution of Eq.~(\ref{g1_6}) for a set of parameters 
$\{ \eta,\mu \}$, then $-E$ is a solution for the set $\{ -\eta,-\mu \}$.
Additionally, given two solutions $E_3$ and $E$ of Eq.~(\ref{g1_6}), it
can be shown that
\begin{equation}
E_3^3 - E^3 + \alpha(E_3^2 - E^2) + \beta (E_3-E)=0.
\label{g2_1}
\end{equation}

If $E_3 \ne E$, the polynomial on the left-hand side can be
simplified, and we find that
\begin{equation}
E^2 + E(E_3 + \alpha) + E_3^2 + \alpha E_3 + \beta = 0,
\label{g1_8}
\end{equation}
so that the respective solutions provide the unaccounted energies
\begin{equation}
\small{
E_1 = -\frac{ E_3 + \alpha + \sqrt{(E_3+\alpha)^2 - 4 (E_3^2 + \alpha E_3 + \beta)}  }{2},
}
\label{g1_9}
\end{equation}
and similarly for $E_2$. Absolute periodicity results when the
energy ratios adhere to the forms
\begin{equation}
x = \frac{E_1}{E_3} = \frac{n_1}{n_3}, \text{ } y = \frac{E_2}{E_3} = \frac{n_2}{n_3}.
\label{g1_10}
\end{equation}

One may ask whether, given a set of parameters $\eta$ and $\mu$, the
system would display periodicity. This however might not be a
convenient approach, since in any case we can find integers $n_1$,
$n_2$ and $n_3$ showing ratios as close to $x$ and $y$ as we
want. This is also true for any reasonable quantum closed
system. Instead, we propose an approach in which, given a pair of
ratios, we ask if a set of parameters yielding those
ratios exits. Following this idea we proceed as follows.

\subsection{Standard Procedure}

Let us then consider $x$ and $y$ as known variables. We divide Eq.~(\ref{g1_9})
by $E_3$ and find $x$. The variable $y$ is obtained in a similar way. By means of
algebraic operations we can express the unknown variables in terms of $\alpha$:
\begin{eqnarray}
& E_3 = -\alpha/(1+x+y), \label{g1_11a} & \\
& \beta = p \alpha^2 , \label{g1_11b} & \\
& \gamma = q \alpha^3,  \label{g1_11c} &
\end{eqnarray}
so that $p$ and $q$ are given by 
\begin{eqnarray}
& p = (x+y+xy)/(1+x+y)^2, \label{g1_12a} & \\
& q = xy/(1+x+y)^3. \label{g1_12b} & 
\end{eqnarray}

From Eqs.~(\ref{g1_7a})~and (\ref{g1_7c}) we find that
\begin{equation}
\gamma = \frac{4}{3}(\eta-\alpha).
\label{g1_13}
\end{equation}

Similarly, using (\ref{g1_13}) in combination with (\ref{g1_11c})
leads to
\begin{equation}
\eta = \alpha \left( 1 + \frac{3q}{4}\alpha^2 \right).
\label{g1_14}
\end{equation}

Insertion of Eq.~(\ref{g1_14})~in (\ref{g1_7a}) then yields
\begin{equation}
\mu = 2\alpha \left( 1 + \frac{5q}{8}\alpha^2 \right).
\label{g1_15}
\end{equation}

The combination of Eqs.~(\ref{g1_7b}), (\ref{g1_11b}), (\ref{g1_14}),
and (\ref{g1_15}) leads to the characteristic equation
\begin{equation}
q^2 A^3+ 2 q A^2 +  4 p A + 16 =0,
\label{g1_16}
\end{equation}
where $A=\alpha^2$. In this way, given a pair of values $x$ and $y$
we find $p$ and $q$ from Eqs.~(\ref{g1_12a})~and (\ref{g1_12b}) and
introduce them in Eq.~(\ref{g1_16}). From the solutions we find
$\alpha$ and therefore the $\eta$ and $\mu$ yielding the 
energy ratios. Strictly speaking, we have 6 solutions, but they come
in pairs giving energies of opposite signs. In order for a solution
to be physically acceptable we demand $\eta$ and $\mu$ to
be real. Let us next discuss particular cases to which the previous method
does not apply.

\begin{figure}
\begin{center}
\includegraphics[width=0.32\textwidth,angle=-90]{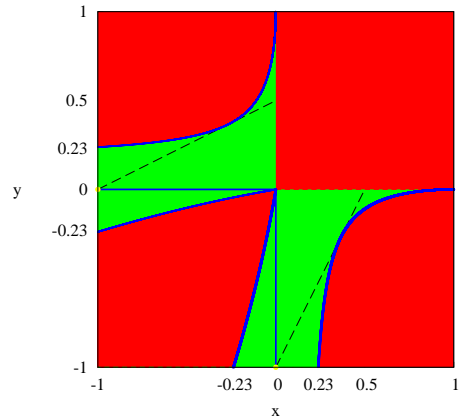}
\caption{(Color online) Coordinate map indicating the number of set of
  parameters $\{\eta,\mu\}$ delivering energy ratios $\{x,y\}$. Red
  (dark gray) region: 0 solutions. Green (light gray) region: 4
  solutions. Blue (continuous) lines: 2 solutions. Yellow dots
  ($\{x=-1,y=0\}$ and $\{x=0,y=-1\}$): 1 solution.  Black (dashed)
  lines highlight the pairs for which there are solutions with
  $\eta=0$.  In these cases, given a couple $\{x,y\}$ there are two
  solutions with $\eta=0$ and two solutions with $\eta\ne0$. The
  mirror symmetry around $y=-x$ derives from the fact that $E_1$ and
  $E_2$ can be swapped in Eqs.~(\ref{g1_10}). Here we require $|E_1|
  \leqslant |E_3|$ and $|E_2| \leqslant |E_3|$. Along the edges of the
  square where $x$ or $y$ is $-1$, there exists an ambiguity in the
  choice of $E_3$ because two eigenenergies display the same absolute
  value, but their signs are opposite. This implies that the ratio
  that is different from $-1$ can either be positive or negative.  As a
  consequence, the borders remain identified: $\{x=-1,y\} \sim
  \{x=-1,-y\}$ and $\{y=-1,x\} \sim \{y=-1,-x\}$. }
\label{fig1}
\end{center}
\end{figure}

\subsection{Particular Cases}

{\it Case A}. $\alpha = x + y + 1 = 0$. \\

For a pair of values $x$ and $y$ the parameter $\eta$
is a solution of the polynomial equation (see appendix A)
\begin{equation}
\kappa^2\eta^6 + 27 \kappa^2\eta^4 + \left( 243 \kappa^2+ \frac{81}{4 } \right) \eta^2 + 729 \kappa^2= 0,
\label{g1_17}
\end{equation}
where we have introduced
\begin{equation}
\kappa^2 = -\frac{4}{3+(x-y)^2} \left( 1 - \frac{4}{3+(x-y)^2} \right).
\label{g1_18}
\end{equation}

Moreover, $\mu$ can be found from Eq.~(\ref{g1_7a}). \\

{\it Case B}. $x=1$ or $y=1$ (excluding $\{x=1,y=0\}$, $\{x=0,y=1\}$ and $\{x=y=1\}$). \\

In order to avoid division by zero in Eq.~(\ref{g2_1}) we modify the
variables in the following way
\begin{equation}
\text{if } x=1 \Rightarrow x'=y'=\frac{1}{y}, 
\label{g1_19}
\end{equation}
and analogously when $y=1$. The new variables then admit the 
standard procedure. \\

{\it Case C}. $\{x=0,y=1\}$ or  $\{x=1,y=0\}$. \\

Then $q=\gamma=0,\beta=-4$ and $\alpha=\eta$. Non-vanishing
energy values satisfy $E^2+\eta E -4 =0$, and 
$x\text{ (or $y$)}=1$ is a solution only if $\eta=\pm4i$. \\

{\it Case D}. $\{x=y=1\}$. \\

The spectrum of the Hamiltonian is threefold
degenerate. Hence the eigenvalue equation is of the
form $(E_3-E)^3=0$. Comparing with Eq.~(\ref{g1_6})
we infer that $-3E=\alpha,3E^2=\beta$ and $-E^3=\gamma$. 
It then follows that $E^6+6E^4+12E^2+16=0$, which has no real
solutions, and therefore no real-valued $\eta$ and
$\mu$ yield a fully degenerate spectrum. 

Fig. \ref{fig1} shows a map classifying the
coordinate space according to the number of valid 
solutions encountered for every pair of ratios $\{x,y\}$.
The maximum number of solutions is 4, usually coming
from 2 real solutions of Eq.~(\ref{g1_16}). 
Along the negative side of the axes the Hamiltonian
becomes reducible with $\mu=2\eta$; hence only two
solutions are possible.  The two instances with one solution correspond to
$\eta=\mu=0$.

\subsection{Recurrences}

\begin{figure}
\begin{center}
\includegraphics[width=0.32\textwidth,angle=-90]{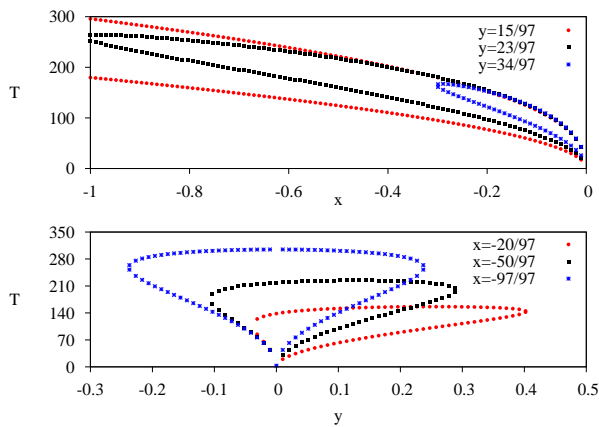}
\caption{(Color online) Top: Absolute period (arbitrary units) as a function of
  $x$, with $y$ kept constant. Bottom: Absolute period as a function of $y$, with
  $x$ kept constant. In both cases the grid slice is $1/97$.  As it
  can be seen, the smallest values of $T$ are found close to the
  origins.}
\label{fig2}
\end{center}
\end{figure}

The condition for absolute periodicity is given in Eq.~(\ref{g2_3}) while
partial periodicity can be determined from
\begin{equation}
E_1 \tau = \phi, \text{ } E_2 \tau = \phi + 2\pi N_2, \text{ } E_3 \tau = \phi + 2\pi N_3,
\label{g3_1}
\end{equation}
in such a way that $N_2$ and $N_3$ are both integers. If $T=N\tau$,
with $N$ an integer, we can infer from Eqs.~(\ref{g2_4})~and (\ref{g3_1}) that
\begin{equation}
\phi = \frac{2\pi n_1}{N}, \text{ } N_2 = \frac{n_2-n_1}{N}, \text{ } N_3 = \frac{n_3-n_1}{N}.
\label{g3_2}
\end{equation}

In order to find $T$ and $\tau$, we first determine $n_1$, $n_2$ and 
$n_3$ from the rationals $x$ and $y$ in such a way that there is no 
common divisor greater than $1$ among the three generating integers. 
Simultaneously, $x$ and $y$ are used to find the
corresponding eigenenergies following the previously discussed method. 
We can then choose any of the identities
in Eqs.~(\ref{g2_4}) (in particular we choose $E_3 T = 2 \pi n_3$) to get 
$T$ using one of the eigenenergies. Finally, the integer $N$ results as 
the greatest common divider of $n_2-n_1$ and $n_3-n_1$. As it can be seen, 
large integers $n_3$ are more likely to yield large $T$ and
$\tau$.

\begin{figure}
\begin{center}
\includegraphics[width=0.32\textwidth,angle=-90]{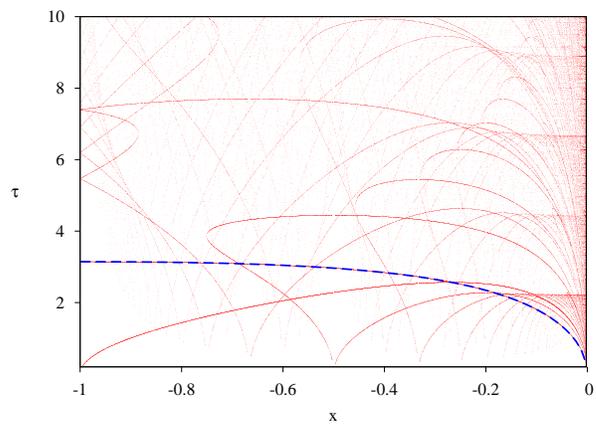}
\caption{(Color online) Partial period (arbitrary units) as a function
  of $x$ for $M=2$ and $M=1$ [Eq.~(\ref{g2_9}) - dashed blue line]. We
  have intentionally restrained the vertical axis to
  $0<\tau<10$. Interacting Hamiltonians can display longer $\tau$ than
  non-interacting Hamiltonians.  For this particular graph we used
  grid slices ranging from $1$ to $1/500$ in order to scan the $xy$
  square of Fig \ref{fig1} (see appendix B).  Although finer
  grids would yield denser graphs, these graphs would not be
  very different from the figure.}
\label{fig3}
\end{center}
\end{figure}

\begin{figure}
\begin{center}
\includegraphics[width=0.32\textwidth,angle=-90]{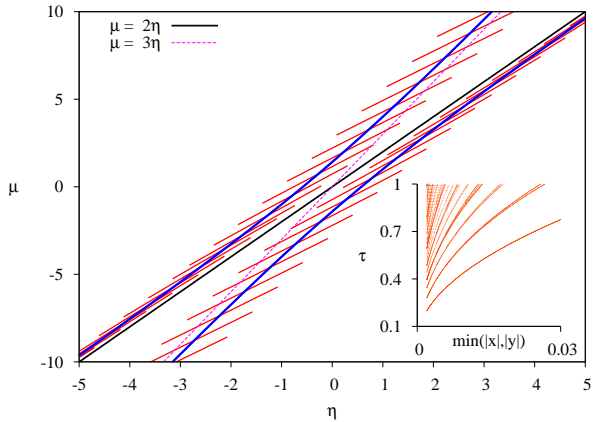}
\caption{(Color online) Blue lines: set of parameters for which the
  period is stable under small changes in the direction indicated by
  the transverse red lines. Inset: $\tau$ (arbitrary units) as a
  function of the smallest of $|x|$ and $|y|$. It follows that as
  $\tau$ goes to zero one of the eigenenergies becomes small in
  comparison to the others.}
\label{fig6}
\end{center}
\end{figure}

Figure~\ref{fig2} presents sample plots of $T$ as a function of $x$
and $y$.  A complete depiction would be quite more intricate. We point
out that $T$ decreases as $x$ and $y$ approach zero. This behavior
appears to be generic, laying minimum values of $T$ around (or in) the
origins. Following $T$ across the line $\{x=-1/n,y=1/n\}$ for
$n=5,6,7,\ldots$, we find that it goes down asymptotically toward $T=\pi
\sqrt{2}$ as $n\rightarrow \infty$. The same minimum value can be
analytically found at $\{x=-1/2,y=0\}$ and its equivalent. Both
instances suggest a relation between the minimum $T$ and the Hamiltonian
matrix being or becoming singular.

Fig. \ref{fig3} depicts the intricate relation between $\tau$ and $x$
obtained by testing energy ratios over the $xy$ square of Fig. \ref{fig1},
as explained in appendix B.
Cooperative systems may display longer partial periods than non-interacting
systems; therefore, periodicity can help identify interacting phases.
This behavior is most likely due to the capacity of many-body systems
to develop complex dynamics on account of the increased number of effective
states. Also of interest is the chaotic
aspect of Fig. \ref{fig3}, indicating that small intervals in $xy$ space do not 
necessarily map onto small intervals in $\tau$ space. This lack 
of continuity, which also characterizes $T$, occurs because  
neighboring values of $x$ and $y$ do not always correspond to 
neighboring values of $n_1$, $n_2$ and $n_3$. 
For instance, $\{x=1/3,y=2/3\}$ is
not far from $\{x=97/300,y=201/300\}$, but the sets of integers that 
generate each pair are widely different. Recall that the period depends
directly on the integers. In addition, there can be
common dividers between $n_2-n_1$ and $n_3-n_1$ and this might further affect 
the continuity of $\tau$. Finally, it can be seen from the inset of
Fig. \ref{fig6} that, when $\tau$ becomes small, at least one of the 
ratios $x$ or $y$ approaches zero, once again suggesting that in a Hamiltonian 
displaying small periodicity one of the eigenenergies is much smaller 
than the others.

\section{Stability}

As introduced here, periodicity, either total or partial, is a
characteristic of the Hamiltonian~\cite{note3}.  This
encourages us to ask whether one can change the Hamiltonian parameters
without affecting the period. This happens for one particle
when $\mu=0$, as can be seen from Eq.~(\ref{g2_9}):

\begin{equation}
\left( \frac{d\tau}{d\mu} \right )_{\mu=0} = 0.
\label{g4_2}
\end{equation}

For two particles, we can start out by arguing that the period is stable
whenever both $x$ and $y$ are stable under changes of the
parameters. Moreover, we can consider the energy ratios as functions
of the parameters, $x(\eta,\mu)$ and $y(\eta,\mu)$, in such a way that
maximum variations occur in the direction of the function gradients
and zero variations take place in the directions perpendicular to the
gradients. In this way, for any set of parameters $\eta$ and $\mu$ one
can always identify a direction of change of the parameters along
which {\it one} of the ratios is stable. It then follows that in order
for $T$ to be stable the directions of zero change of both ratios must
coincide, i.e., the gradients must be parallel:
\begin{equation}
\vec{\nabla} x + \lambda \vec{\nabla} y =0,
\label{g4_1}
\end{equation}
where $\vec{\nabla} x = (\partial x/\partial \eta) \vec{\eta} + (\partial
x/\partial \mu) \vec{\mu}$, and similarly for $\vec{\nabla} y$. The vectors
$\vec{\eta}$ and $\vec{\mu}$ are unitary vectors in parameter space
and $\lambda$ is a real number. This problem is equivalent to finding
the extreme values of $x(\eta,\mu)$ subject to the condition
$y(\eta,\mu)$=constant, or the other way around. In this context
$\lambda$ takes the role of a Lagrange multiplier and the analogy
applies as long as the involved functions are smooth.  The extremes of
$x$ can be worked out from Fig. \ref{fig1}. It is found, however, that
not all extreme values imply parallel gradients. For instance, along
the edges, where $x$ or $y$ is $-1$, the identification of borders on
each side of the axis generates a discontinuity in the first
derivative of the ratios and the analogy with the Lagrange method
does not apply. Likewise, for the extremes located around
the neighborhood of the axes at least one of the parameters diverge
toward infinity and there is no solution of Eq.~(\ref{g1_16}) along the
positive side of the axes.  In general, we find the gradients to be
parallel only along the blue (continuous) lines of Fig.~\ref{fig1} located
between green and red regions.

Figure~\ref{fig6} shows curves indicating the parameters as well as the
directions of change corresponding to zero variation of the
period. Far from the origin the curves approach (but do not seem to
touch) straight lines given by simple expressions. In one case the
direction of zero change aligns with the direction of the curve as we
get away from the origin. In the other case the direction of zero
change becomes constant with an angle of inclination $\theta$
satisfying $\tan \theta \approx 1/\sqrt 2$.

\begin{figure}
\begin{center}
\includegraphics[width=0.32\textwidth,angle=-90]{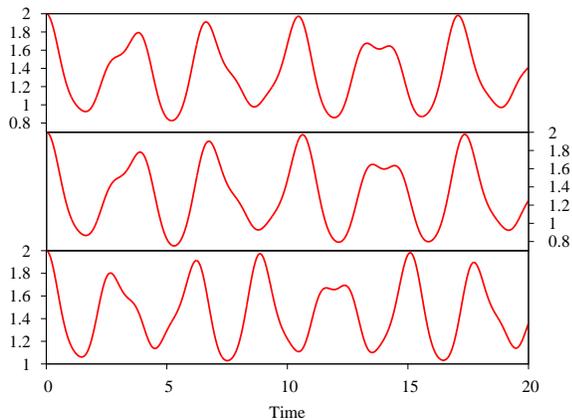}
\caption{(Color online) Mean number of particles $\langle \hat
  a_1^{\dagger} \hat a_1 \rangle$ as a function of time (arbitrary
  units) for three slightly different cases. Top panel: $\eta =
  1.878615$ and $\mu=2.939702$.  Middle panel: $\eta = 1.778615$,
  $\mu=2.775950$. Bottom panel: $\eta=1.878615$ and $\mu=2.714882$.  The
  middle panel corresponds to a set of parameters in the curve of
  stability of Fig.~\ref{fig6}. The top and bottom plots show the
  evolution of the system when such parameters are shifted in the
  direction of zero change and in the direction of the gradient,
  respectively.}
\label{fig7}
\end{center}
\end{figure}

As can be seen from Fig. \ref{fig7}, when the parameters are shifted in
the direction of zero change the dynamics of corresponding systems look 
similar. Conversely, when the change of parameters occurs in the
direction of the gradient, the dynamics soon diverge. This behavior
is consistent with our study and shows that in certain cases the
period alone characterizes the global profile of the system evolution.

\subsection{Application to quantum optics: stability condition for 
coherent population trapping}

Besides applicable to systems described by the
Hamiltonian~(\ref{g1_1}), our results can be extrapolated to models
with analogous Hamiltonians. Let us consider a three-level quantum
system interacting with two lasers. We focus on the phenomenon known
as coherent population trapping (CPT) \cite{Scully}. In the
semiclassical approach the Hamiltonian can be written in the form
{\small
\begin{equation}
\sum_{k=0}^2 E_k |k\rangle \langle k|  
-\Omega \left ( e^{-i\phi_0-i\omega_0 t}  |1\rangle \langle 0| + e^{-i\phi_2-i\omega_2 t}  |1\rangle \langle 2| + h.c. \right).
\label{leq:1}
\end{equation}
}

Following the notation in Ref.~\cite{Scully}, we define $\Omega
e^{-i\phi_0}$ and $\Omega e^{-i\phi_2}$ as the complex Rabi
laser frequencies. We have already assumed that both lasers
have the same intensity. $E_0$,$E_1$, and $E_2$ are the eigenenergies
of the unperturbed levels, $|0\rangle$, $|1\rangle$ and $|2\rangle$ respectively. We 
propose a ladder scheme where $E_2>E_1>E_0$. Similarly, $\omega_0$ and $\omega_2$ are the laser
frequencies. CPT means that, as the system evolves, $|1\rangle$ remains unpopulated as a
result of quantum interference. The characteristic state of CPT is known as a dark state
and is associated to a zero eigenvalue. Measuring
energy in units of $\Omega / \sqrt{2}$ Hamiltonian (\ref{leq:1}) reads

{\small
\begin{equation}
\sum_{k=0}^2 E_k' |k\rangle \langle k|  
-\sqrt{2} \left ( e^{-i\phi_0-i\omega_0 t}  |1\rangle \langle 0| + e^{-i\phi_2-i\omega_2 t}  |1\rangle \langle 2| + h.c. \right).
\label{leq:22}
\end{equation}
}

Likewise, introducing the following time-dependent unitary transformation

\begin{equation}
\hat U = -e^{-i\phi_0 - i \omega_0 t} |0\rangle \langle 0| - e^{-i\phi_2 - i \omega_2 t} |2\rangle \langle 2| + |1\rangle \langle 1|,
\label{leq:2}
\end{equation}
we find that the  transformed state $|\psi_I \rangle = \hat U |\psi \rangle$  evolves as
\begin{equation}
i\frac{d|\psi_I \rangle}{dt} = \left( \hat{H}_I + i \left( \frac{d\hat{U}}{dt} \right) \hat{U}^{-1} \right) |\psi_I \rangle = \hat{H}_D|\psi_I \rangle,
\label{leq:3}
\end{equation}
so that, $\hat {H}_I = \hat U \hat {H} \hat {U}^{-1}$ with $\hat{H}$ given
by Eq.~(\ref{leq:22}). Evolution is now determined by the time-independent
Hamiltonian $\hat {H}_D$. Making use of the completeness relation
to write $|2\rangle \langle 2| = 1-|0\rangle \langle
0|-|1\rangle \langle 1|$, $\hat {H}_D$ becomes
 
\begin{equation}
\left (
\begin{array}{ccc}
\omega_0-\omega_2-{\Omega}(E_2' - E_0') & \sqrt{2} & 0 \\
\sqrt{2} & -\omega_2-(E_2' - E_1') & \sqrt{2} \\
0 & \sqrt{2} & 0
\end{array}
\right ).
\label{leq:4}
\end{equation}

Direct comparison with (\ref{g1_5}) then establishes that
\begin{eqnarray}
4\eta - 2 \mu = \omega_0-\omega_2-\Delta_0-\Delta_2, \label{leq:5} \\
\eta - \mu = -\omega_2-\Delta_2, \label{leq:6}
\end{eqnarray}

where we have introduced $\Delta_0 = E_1'-E_0'$ and $\Delta_2 = E_2'-E_1'$.
In the ratio diagram of Fig.~\ref{fig1}, the regions
describing a Hamiltonian with one vanishing eigenvalue
correspond to the x- and y-axis. These have no intersection
with the regions of stability for the total period, 
which are located between the green and red regions. 
The partial period is however stable at $\{x=-1,y=0\}$. 
These ratios correspond to $\eta = \mu = 0$. Replacing
such values in Eqs.~(\ref{leq:5})~and (\ref{leq:6}) we find that
\begin{eqnarray}
\omega_0 = \Delta_0,\label{leq:7} \\
\omega_2 = -\Delta_2.\label{leq:8}   
\end{eqnarray}

These identities confirm that CPT arises when the
frequencies of the lasers coincide with the energy difference
between levels. Since CPT is a dynamical phenomenon, it is 
possible that deviations of the parameters from Eqs.~(\ref{leq:7})~and (\ref{leq:8})
reduce its efficiency. Nevertheless,
since $\tau$ is stable under variations of $\mu$ around $\mu=0$
in the single-particle case, which corresponds to $\eta=0$, we can derive
parallel stability conditions for CTP from Eqs.~(\ref{leq:5})~and (\ref{leq:6})

\begin{eqnarray}
- 2 d\mu = d\omega_0 - d\omega_2 - d\Delta_0 - d\Delta_2, \label{leq:9}\\
 - d\mu = -d\omega_2 - d\Delta_2. \label{leq:10}
\end{eqnarray}

Finally, replacing Eq.~(\ref{leq:10}) in Eq.~(\ref{leq:9}) yields

\begin{equation}
d\Delta_2 - d\Delta_0 + d\omega_2 + d\omega_0 = 0.
\label{leq:12}
\end{equation}

It can then be inferred that given a situation
where CPT is dominant, i.e., where Eqs.~(\ref{leq:7})
and (\ref{leq:8}) both hold, the dynamics is stable
under small deviations of the parameters as long as
such deviations are correlated as in Eq.~(\ref{leq:12}).

\section{Conclusions}
We studied the time periodicity as well as the stability
properties of interacting bosonic systems. The difference
between the usual notion of periodicity, in which the ratios of the
Hamiltonian energies become commensurate, and partial periodicity, in
which the state recurs up to a phase factor, was stressed. 
Both forms of periodicity
were explicitly established for one and two particles in a model
described by two bosonic modes.  The results suggest a connection
between minimum periodicity and the fact that one of the eigenvalues
becomes small in comparison to the other eigenvalues, especially in
the two-particle case \cite{note1}. Similarly, we pointed out that the
stability of the period depends not only on the parameters, but
also on the direction of change of such parameters.  To emphasize this
fact, a diagram showing the set of stability parameters as well as the
directions of change for which the period stays 
constant was presented. We found that these results are consistent with 
simulations of the dynamics and apply the formalism to find
stability conditions of CPT.  

In should be noted that the assumption $T=N\tau$ made for the two-particle 
case does not cover the whole set of possilibities of periodicity. It may be
possible to relax this assumption and derive $\tau$ using only
Eq.~(\ref{g3_1}). This would in principle lead to a richer stability
diagram including parameters for which $\tau$ is stable, but $T$ is
not, e.g., $\eta=\mu=0$.  Similarly, we feel that our method
can be extended to more complex Hamiltonians.

\appendix

\section{Appendix A: Derivation of Eq.~(\ref{g1_17}).}

When $\alpha=0$ the solutions of Eq.~(\ref{g1_8}) can be written in
the form
\begin{equation}
E_1 = \frac{-E_3-\sqrt{-3 E_3^2-4\beta}}{2}, \hspace{0.2cm} E_2 = \frac{-E_3+\sqrt{-3 E_3^2-4\beta}}{2}.
\label{a1_1}
\end{equation}

It then follows that
\begin{equation}
E_3 = \sqrt{\frac{-4\beta}{3+(x-y)^2}}.
\label{a1_2}
\end{equation}

Substituting Eq.~(\ref{a1_2}) in Eq.~(\ref{g1_6}) with $\alpha=0$, we have the
equality
\begin{equation}
\kappa \beta^{3/2} = \gamma,
\label{a1_3}
\end{equation}
where $\kappa$ is given by Eq.~(\ref{g1_18}). Likewise, from
Eq.~(\ref{g1_7a}) $\mu =\frac{5}{3}\eta$. 
Replacing this $\mu$ in Eqs. (\ref{g1_7b}) and (\ref{g1_7c}) gives

\begin{equation}
\beta = -4 \left( \frac{\eta^2}{9}+1 \right), \text{ } \gamma = \frac{4}{3} \eta.
\label{a1_4}
\end{equation}

Substitution in Eq.~(\ref{a1_3}) and some algebra then leads to Eq.~(\ref{g1_17}).

\section{Appendix B: Generation of commensurate ratios.}

In order to generate pairs of ratios compatible with the coordinate
square in Fig.~\ref{fig1}, we first choose an integer $n_3=1,2,3,...$,
which is also the inverse of the grid slice.  For a given $n_3$, we
generate integers in the range $-n_3<n_1<-1$ and $n_1<n_2<n_3-1$. 
The values of $n_3$ are
inserted into a computer routine in increasing order, starting with
$n_3=1$.  Subsequently, $n_1$ and $n_2$ are introduced.  These
integers determine $x$ and $y$, which are in turn used to check for
Hamiltonians with corresponding eigenenergies.  Since the coordinate
square is symmetric, we only have to scan the section $x<0$. If for the
integers in a
given triplet $\{n_1,n_2,n_3\}$ we find a maximum common divider
greater than 1, the triplet is discarded. Notice should be taken that in this
procedure the coordinate square is scanned using several superimposed
grids, avoiding repetition of $\{x,y\}$ pairs.


\begin{thebibliography}{}
\bibitem{note2} As $J$ usually depends on the details of the
system, we avoid making reference to it and use arbitrary units
for our time scales.
\bibitem{note3} This statement is valid as long as the initial state
is not an eigenstate of the Hamiltonian.
\bibitem{note1} The only exception takes place for one particle
when $\mu \rightarrow \infty$ since according to Eqs. (\ref{g2_5})
and (\ref{g2_6}) $T\rightarrow \infty$.
\end{thebibliography}
\end{document}